\documentclass[fp,twocolumn,letter]{jpsj3}
\usepackage{color}
\usepackage{bm}
\usepackage{mathrsfs}
\usepackage{braket}

\newcommand{\mr}[1]{{\mathrm{#1}}}
\newcommand{\non}{{\nonumber}}

\title{Typical Purification Reproducing the Time Evolution of an Open Quantum System}
\author{Koki Inoue and Yoshiyuki Fukumoto}
\inst{Department of Physics, Faculty of Science and Technology, Tokyo University of Science, Noda, Chiba 278-8510, Japan
}
\recdate{\hspace{4cm}}

\abst{
　\\
It is known that an arbitrary quantum state can always be expressed in a pure state by preparing an appropriate ancilla system, which is called purification. 
The purification can always be performed when considering only a quantity not dependent on the quantum state of the ancilla system. Otherwise, the possibility of purification cannot be elucidated. 
In this study, we focus on the cases where a quantum system S and the ancilla system interact with each other, and investigate whether purification, which correctly reproduces the time evolution of the system S, is possible. Accordingly, when the ancilla system is a macrosystem, it can be shown that this purification is possible for a specific initial state.
}

\begin{document}
\maketitle

A quantum system S could be in a mixed state, which is impossible to describe with a single ket vector.
However, even such a mixed state can always be expressed by using a pure state in a composite system of the system S and an ancilla system E.
That is, let $\hat \sigma_{\mr{S}}:=\sum_{\mu}q_{\mu} \ket{\mu}_{\mr{S\,S}}\hspace{-0.2em}\bra{\mu}$ be a density operator of S, 
where $\{\ket{\mu}_{\mr{S}}\}_{\mu}$ is S's orthonormal basis and $\{q_{\mu}\}_{\mu}$ is the real number sequence satisfying $q_{\mu}\geq 0$ and $\sum_{\mu} q_{\mu}=1$; then
we can construct a pure state of the composite system $\mr{S}+\mr{E}$,
\begin{align}
\label{purification}
\ket{\psi_{\mr{S+E}}}:=\sum_{\mu} \sqrt{q_{\mu}}\ket{\mu}_{\mr{S}}\otimes\ket{\mu}_\mr{E},
\end{align}
with an orthonormal system $\{ \ket{\mu}_{\mr{E}} \}_{\mu}$ for E.
Then, the mixed state $\hat \sigma_{\mr{S}}$ is obtained as the reduced state of S for $\ket{\psi_{\mr{S+E}}}$ in Eq. (\ref{purification}).
This operation for constructing a pure state, such as $\ket{\psi_{\mr{S+E}}}$, is called ``purification'' and is mainly used in the context of quantum information \cite{quantuminfo}.
Usually, in purification, the reduced state of S is physically meaningful but E is not, and thus we can choose the Hilbert space of E and E's ket vectors arbitrarily.

In this letter, we consider the opposite case where the ancilla system E is physically meaningful and affects the state of the system S.
Specifically, we assume the interaction between S and E to be switched on at time $t=0$ to study the time evolution of S at $t>0$, with
the actual initial state being a product state like $\hat \sigma_{\mr{S}}\otimes \hat \sigma_{\mr{E}}$,
where $\hat \sigma_{\mr{S}}$ and  $\hat \sigma_{\mr{E}}$ denote the density operators of S and E, respectively.
Obviously, the time evolution of ${\rm S} + {\rm E}$ is determined by the total Hamiltonian including interactions.
In this case, not only is the above arbitrariness in the purification procedure lost, but also it is not clear whether an arbitrary quantum state of S can always be expressed by using a pure state of ${\rm S} + {\rm E}$.

We define ``purification reproducing the time evolution (PRT)'' as constructing a pure initial state for which the reduced state is equal to the actual reduced state at almost all times of $t=0\sim\tau$.
It is very interesting to study the conditions under which PRT is possible.
In this letter, setting E as a thermal bath B, in order to show the existence of PRT for the composite system ${\rm S} + {\rm B}$, we focus on the typicality of the equilibrium state of the macroscopic system.
The typicality of the equilibrium state refers to the fact that almost all pure states in the appropriate subspace of Hilbert space have the expected value of the macroscopic quantity equal to the equilibrium value\cite{sugita}.
It is plausible that the typical states of B have typical effects on S, and thus we can purify any state of S by using the typical states of B.
Accordingly, we will show that almost all pure states thus obtained are PRTs of S.

Our discussion also relates to the derivation of the quantum master equation, which describes the time evolution of an open system weakly coupled with a thermal bath.
The initial state of the composite system is limited to those having no correlation between S and the thermal bath in the derivation of this equation.
As described later, by applying our results to weakly coupled systems, 
we show that the application range of the quantum master equation is expanded to the initial state having a correlation between S and the thermal bath.


Now we consider the mathematical aspect.
Let $\mathcal{H}_{\mr{S}}$ be a Hilbert space of S and the dimension of $\mathcal{H}_{\mr{S}}$ be $D_{\mr{S}}$, 
and $\mathcal{H}_{\mr{B}}$ be the Hilbert space of B composed of $N_{\mr{B}}$ particles.
We define $\mathcal{E}_{U,\Delta}$ as the subspace of $\mathcal{H}_{\mr{B}}$ spanned by the energy eigenstates in which the energy eigenvalues enter $(U-\Delta,U]$, and define $\hat \rho_{U,\Delta}^{\mr{m.c.}}$ as a density operator of a microcanonical ensemble defined by the energy range $(U-\Delta ,U]$.
We assume that the actual initial state of S$+$B is 
\begin{align}
   \hat \rho_{\mr{S}}^{\mr{R}}(0)\otimes \hat \rho_{U,\Delta}^{\mr{m.c.}},
\end{align}
where $\hat \rho_{\mr{S}}^{\mr{R}}(0)$ is an arbitrary density operator of S.
It can be decomposed into 
\begin{align}
   \hat \rho_{\mr{S}}^{\mr{R}}(0)=\sum_{j=1}^{D_{\mr{S}}}p_j \ket{\,j\,}\bra{\,j\,},
\end{align}
with S's orthonormal basis $\{\ket{\,j\,}\}_{j=1}^{D_{\mr{S}}}$ and the real number sequence $\{p_j\}_{j=1}^{D_{\mr{S}}}$ satisfying $p_j\geq 0$ and $\sum_{j=1}^{D_{\mr{S}}} p_j=1$.
We show that there are always typical pure states of S$+$B, which correctly reproduce the actual time evolution of S.
Assuming that B is consistent with thermodynamics, $D_{\mr{S}}\ll D_{U,\Delta}$ holds at sufficiently large $N_{\mr{B}}$,
where $D_{U,\Delta}$ is the dimension of $\mathcal{E}_{U,\Delta}$.
Let $\{\ket{\psi_j}\}_{j=1}^{D_{\mr{S}}}$ be a random orthonormal system, which is chosen uniformly from $\mathcal{E}_{U,\Delta}$.
That is, $\{\ket{\psi_j}\}_{j=1}^{D_{\mr{S}}}$ obey a probability density function, $P\left(\{\ket{\psi_j}\}_{j=1}^{D_{\mr{S}}}\right)$, which is defined as 
\begin{align}
P\left(\{\ket{\psi_j}\}_{j=1}^{D_{\mr{S}}}\right):=&A\,\prod_{j=1}^{D_{\mr{S}}}\delta\left(\,\braket{\psi_j|\psi_j}-1\,\right)\non \\
&\times\prod_{k > j}^{D_{\mr{S}}}\delta\left(\,\mr{Re}(\braket{\psi_j|\psi_k})\,\right)\delta\left(\,\mr{Im}(\braket{\psi_j|\psi_k})\,\right),
\end{align}
where $A$ is the normalization constant.
Then a pure state of S$+$B,
\begin{align}
   \label{syoki}
   \ket{\psi_{\mr{S+B}}}:=\sum_{j=1}^{D_{\mr{S}}} &\sqrt{p_j} \ket{\,j\,} \otimes \ket{\psi_j},
\end{align}
gives a purification of $\hat \rho_{\mr{S}}^{\mr{R}}(0)$.
Furthermore, since $\{\ket{\psi_j}\}_{j=1}^{D_{\mr{S}}}$ are typical states that are indistinguishable from the microcanonical ensemble with high probability\cite{sugita}, the pure state in Eq. (\ref{syoki}) reproduces the macroscopic state of B.
Let $\hat V(t)$ be a unitary operator that develops the state of S$+$B from time $0$ to time $t$, 
then the reduced state at time $t$ under initial state $\hat \rho_{\mr{S}}^{\mr{R}}(0)\otimes \hat \rho_{U,\Delta}^{\mr{m.c.}}$ is given by
\begin{align}
\label{RR}
   \hat \rho_{\mr{S}}^{\mr{R}}(t)&:=\mr{Tr_B}\left(\hat V(t)\hat \rho_{\mr{S}}^{\mr{R}}(0)\otimes \hat \rho_{U,\Delta}^{\mr{m.c.}}\hat V^{\dagger}(t)\right),
\end{align}
and that under the initial state $\ket{\psi_{\mr{S+B}}}$ is given by
\begin{align}
   \label{RP}
   \hat \rho_{\mr{S}}(t)&:=\mr{Tr_B}\left(\hat V(t)\ket{\psi_{\mr{S+B}}}\bra{\psi_{\mr{S+B}}}\hat V^{\dagger}(t)\right),
\end{align}
where $\mr{Tr_B}(\ \cdot\ )$ is a partial trace of B that maps an operator of S$+$B to an operator of S.
We can show that, for any $t$,
\begin{align}
\label{ave}
\overline{\hat \rho_{\mr{S}}(t)}=\hat \rho_{\mr{S}}^{\mr{R}}(t),
\end{align}
where the overline represents the average related to the random variables in $\{\ket{\psi_j}\}_{j=1}^{D_{\mr{S}}}$.

We introduce the time-development distance of the two states in Eqs. (\ref{RR}) and (\ref{RP}),
\begin{align}
\label{jikankyori}
   \left\| \hat \rho_{\mr{S}}-\hat \rho_{\mr{S}}^{\mr{R}} \right\|_\tau :=\sqrt{\frac{1}{\tau}\int_0^\tau dt\ \left\| \hat \rho_{\mr{S}}(t)-\hat \rho_{\mr{S}}^{\mr{R}}(t) \right\|^2},
\end{align}
where $\left\| \cdots \right\|$ represents the Hilbert Schmidt norm of S.
Denoting the partial trace of S as $\mr{Tr_S}(\ \cdot\ )$,
we define a projection superoperator $\mathcal{\hat P}$ as
\begin{align}
   \mathcal{\hat P}(\ \cdot\ ):=\frac{1}{D_{\mr{S}}}\hat I\otimes \mr{Tr_S}(\ \cdot\ ).
\end{align}
Then, the random average $\overline{\left\| \hat \rho_{\mr{S}}(t)-\hat \rho_{\mr{S}}^{\mr{R}}(t) \right\|^2}$ can be evaluated as follows:
\begin{align}
\label{ine}
&\overline{\left\| \hat \rho_{\mr{S}}(t)-\hat \rho_{\mr{S}}^{\mr{R}}(t) \right\|^2}
=\overline{\mr{Tr}\left( \hat \rho_{\mr{S}}(t)\right)^2}-\mr{Tr}\left(\hat \rho_{\mr{S}}^{\mr{R}}(t)\right)^2
\non \\
&<\ D_{\mr{S}}\mr{Tr_{S+B}}\left(\mathcal{\hat P}\left(\hat V(t)\hat \rho_{\mr{S}}^{\mr{R}}(0)\otimes \hat \rho_{U,\Delta}^{\mr{m.c.}}\hat V^{\dagger}(t)\right)\right)^2+\frac{\eta}{D_{U,\Delta}}\non \\
&\leq\ D_{\mr{S}}\mr{Tr_{S+B}}\left(\hat V(t)\hat \rho_{\mr{S}}^{\mr{R}}(0)\otimes \hat \rho_{U,\Delta}^{\mr{m.c.}}\hat V^{\dagger}(t)\right)^2+\frac{\eta}{D_{U,\Delta}}\non \\
&\leq\ (D_{\mr{S}}+\eta)/D_{U,\Delta},
\end{align}
where $\mr{Tr_{S+B}}(\ \cdot\ )$ represents the trace of S$+$B and
\begin{align}
   \eta := \frac{D_{U,\Delta}+D_{\mr{S}}}{D_{U,\Delta}^2-1}.
\end{align}
The parameter $\eta$ comes from the fact that $\{\ket{\psi_j}\}_{j=1}^{D_{\mr{S}}}$ is an orthonormal system, and as $N_{\mr{B}}$ increases, $\eta$ approaches $0$ exponentially.
By using Eq. (\ref{ine}), for arbitrary $\varepsilon >0$, we obtain two inequalities,
\begin{align}
\label{Pr1}
&\mr{Pr}\left(\left\|\hat \rho_{\mr{S}}(t)-\hat \rho_{\mr{S}}^{\mr{R}}(t)\,\right\|\geq \varepsilon\right)<\frac{1}{\varepsilon^2}\frac{D_{\mr{S}}+\eta}{D_{U,\Delta}}, \\
\label{Pr2}
&\mr{Pr}\left(\left\|\hat \rho_{\mr{S}}-\hat \rho_{\mr{S}}^{\mr{R}}\,\right\|_\tau\geq \varepsilon\right)<\frac{1}{\varepsilon^2}\frac{D_{\mr{S}}+\eta}{D_{U,\Delta}},
\end{align}
where $\mr{Pr}(x)$ represents the probability of satisfying the condition $x$.
Equations (\ref{Pr1}) and (\ref{Pr2}) hold for any $t$ and any $\tau$.
The right-hand sides of these equations rapidly approach 1, as $D_{U,\Delta}$ increases exponentially with increasing $N_{\mr{B}}$. 
Therefore, it can be seen that the typical initial state, $\ket{\psi_{\mr{S+B}}}$, is the PRT of $\hat \rho_{\mr{S}}^{\mr{R}}(0)$ with $\hat \rho_{U,\Delta}^{\mr{m.c.}}$ as the state of the ancilla system.

Our results, viz., Eqs. (\ref{ave}), (\ref{Pr1}), and (\ref{Pr2}), can be substantially generalized to other states of B.
Let $\hat \rho_{\mr{ens}}$ be a density operator of B satisfying $\mr{Tr}\,\hat \rho_{\mr{ens}}=1$.
If the ensemble $\hat \rho_{\mr{ens}}$ satisfies 
\begin{align}
D_{\mr{S}}^{2}\ll 1/\mr{Tr}\left(\hat \rho_{\mr{ens}}\right)^2
\end{align}
at sufficiently large $N_{\mr{B}}$, the same result can be obtained by using $\hat \rho_{\mr{ens}}$ instead of $\hat \rho_{U,\Delta}^{\mr{m.c.}}$. 
In that case, instead of $\{\ket{\psi_j}\}_{j=1}^{D_{\mr{S}}}$, we use typical pure states,\cite{sugiura1,sugiura2,sugiura3,gene}
\begin{align}
\ket{\psi_j'}:=\sum_{\nu}\frac{1}{\sqrt{2}}\left(x^{(j)}_{\nu}+i y^{(j)}_{\nu}\right)\sqrt{\hat \rho_{\mr{ens}}}\ket{\nu},
\end{align}
where $x^{(j)}_{\nu}$ and $y^{(j)}_{\nu}$ are random numbers from a standard normal distribution, $\{\ket{\nu}\}_{\nu}$ is an orthonormal basis of B, and $i$ is the imaginary unit.
We can show that the independently prepared $D_{\mr{S}}$ vectors, $\{\ket{\psi_j'}\}_{j=1}^{D_{\mr{S}}}$, form an almost orthonormal system, i.e., for an arbitrary $\varepsilon>0$
\begin{align}
\mr{Pr}\left(\underset{j,k}{\mr{max}}\left\{\,|\hspace{-0.2em}\braket{\psi_j'|\psi_k'}-\delta_{j,k}|\,\right\}\geq \varepsilon\right)\leq \frac{\xi}{\varepsilon^2}\mr{Tr}\left(\hat \rho_{\mr{ens}}\right)^2,
\end{align}
where $\xi:=(D_{\mr{S}}^2+3D_{\mr{S}})/2$.
We can obtain an initial state,
\begin{align}
\ket{\psi_{\mr{S+B}}'}:=\sum_{i=1}^{D_{\mr{S}}}\sqrt{p_j}\ket{j}\otimes\ket{\psi_j'},
\end{align}
which is almost a PRT.
Then, $D_{U,\Delta}$ in Eqs.~(\ref{Pr1}) and (\ref{Pr2}) is replaced by $1/\mr{Tr}\left(\hat \rho_{\mr{ens}}\right)^2$, and $\eta$ is replaced by $0$.
Therefore, it can be seen that PRT can be obtained by purification using the typical states of B.

In particular, the method developed in
Refs.~\citen{sugiura1,sugiura2,sugiura3} can be easily implemented in numerical calculations.
When strong interactions between S and B prevent the use of the quantum master equation, it would be effective to calculate the time evolution of S by using $\ket{\psi_{\mr{S+B}}'}$.
Moreover, when S weakly couples with B, $\hat \rho_{\mr{S}}^{\mr{R}}(t)$ is a solution of the quantum master equation; so Eqs.~(\ref{Pr1}) and (\ref{Pr2}) show that $\hat \rho_{\mr{S}}(t)$ develops according to the quantum master equation.
This implies that the applicable range of the quantum master equation can be extended to states with an initial correlation.

In this study, the purification of quantum states has been discussed from the viewpoint of time evolution of open quantum systems.
First, when the initial state of the system and ancilla system is in the product state, we define PRT as purification that reproduces the time evolution of the system.
We have shown that PRT is always possible if the reciprocal of purity, $1/\mr{Tr}\left(\hat \rho_{\mr{ens}}\right)^2$, is very large compared to $D_{\mr{S}}^2$
by limiting it to the case where the ancilla system is a thermal bath.
This result is established regardless of the Hamiltonian.
Therefore, it can be applied to the numerical calculation of strong coupling systems.
Furthermore, in the case of weak coupling, the application range of the quantum master equation is extended, and the meaning of the hypothesis of the conventional initial correlation is clarified.


{\bf Acknowledgments} This work was partly supported by JSPS KAKENHI Grant Number 17K05519.

\end{document}